\documentclass[twocolumn,english]{revtex4}
\usepackage[T1]{fontenc}
\usepackage[latin9]{inputenc}
\usepackage{color}
\usepackage{amsmath}
\usepackage{graphicx}
\usepackage{amssymb}

\makeatletter
\@ifundefined{textcolor}{}
{%
 \definecolor{BLACK}{gray}{0}
 \definecolor{WHITE}{gray}{1}
 \definecolor{RED}{rgb}{1,0,0}
 \definecolor{GREEN}{rgb}{0,1,0}
 \definecolor{BLUE}{rgb}{0,0,1}
 \definecolor{CYAN}{cmyk}{1,0,0,0}
 \definecolor{MAGENTA}{cmyk}{0,1,0,0}
 \definecolor{YELLOW}{cmyk}{0,0,1,0}
 }

\makeatother

\makeatother

\usepackage{babel}

\makeatother

\usepackage{babel}

\makeatother

\usepackage{babel}

\begin{document}

\title{Optical characterization of topological insulator surface states:
Berry curvature-dependent response}

\author{Pavan Hosur}

\affiliation{Department of Physics, University of California, Berkeley}
\begin{abstract}
We study theoretically the optical response of the surface states
of a topological insulator, especially the generation of helicity-dependent
direct current by circularly polarized light. Interestingly, the dominant
current, due to an interband transition, is controlled by the Berry
curvature of the surface bands. This extends the connection between
photocurrents and Berry curvature beyond the quasiclassical approximation
where it has been shown to hold. Explicit expressions are derived
for the (111) surface of the topological insulator Bi$_{2}$Se$_{3}$
where we find significant helicity dependent photocurrents when the
rotational symmetry of the surface is broken by an in-plane magnetic
field or a strain. Moreover, the dominant current grows linearly with
time until a scattering occurs, which provides a means for determining
the scattering time. \textcolor{black}{The dc spin generated on the
surface is also dominated by a linear-in-time, Berry curvature dependent
contribution.}
\end{abstract}
\maketitle

\section{introduction}

Topological insulators (TIs) are characterized by topologically protected
surface states (SSs). In their simplest incarnation, these correspond
to the dispersion of a single Dirac particle, which cannot be realized
in a purely two dimensional band structure with time reversal invariance.
This dispersion is endowed with the property of spin-momentum locking,
i.e., for each momentum there is a unique spin direction of the electron.
Most of the experimental focus on TIs so far has been towards trying
to directly observe these exotic SSs in real or momentum space, in
tunneling\cite{BiSbSTM} and photoemission\cite{Chen,Hsieh,Xia} experiments,
respectively, and establish their special topological nature. However,
there has so far been a dearth of experiments which study the response
of these materials to external perturbations, such as an external
electromagnetic field.

In order to fill this gap, we study here the response of TI surfaces
to circularly polarized (CP) light. Since photons in CP light have
a well-defined angular momentum, CP light can couple to the spin of
the surface electrons. Then, because of the spin-momentum-locking
feature of the SSs, this coupling can result in dc transport which
is sensitive to the helicity (right- vs left-circular polarization)
of the incident light. This phenomenon is known as the circular photogalvanic
effect (CPGE). In this work, we derive general expressions for the
direct current on a TI surface as a result of the CPGE at normal incidence
within a two-band model and estimate its size for the (111) surface
of Bi$_{2}$Se$_{3}$, an established TI, and find it to be well within
measurable limits. Since bulk Bi$_{2}$Se$_{3}$ has inversion symmetry
and the CPGE, which is a second-order non-linear effect, is forbidden
for inversion symmetric systems, this current can only come from the
surface.

We find, remarkably, that the dominant contribution to the current
is controlled by the \emph{Berry curvature} of the electron bands
and \emph{grows linearly with time}. In practice this growth is cut-off
by a scattering event which resets the current to zero. At the microscopic
level, this part of the current involves the absorption of a photon
to promote an electron from the valence to the conduction band. The
total current contains two other terms - both time-independent - one
again involving an interband transition and the other resulting from
intraband dynamics of electrons. However, for clean samples at low
temperatures, the scattering or relaxation time is expected to be
large, and these contributions will be eclipsed by the linear-in-time
one. Hence, this experiment can also be used to measure the relaxation
time for TI SSs.

\begin{figure}
\begin{centering}
\includegraphics[scale=0.35]{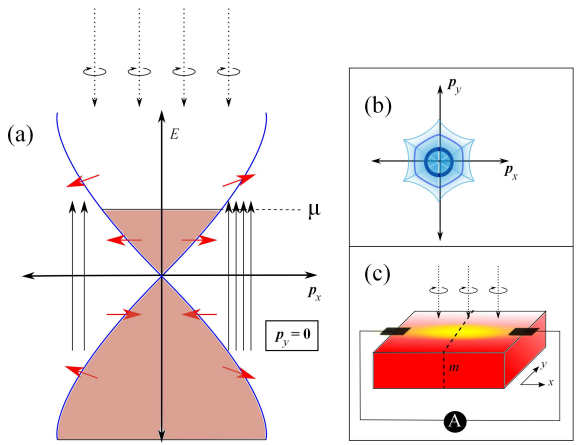}
\par\end{centering}

\caption{(Color online) (a) Schematic illustration of preferential absorption
at one out of two points related by the reflection symmetry about
the $yz$-plane. The short arrows denote the spin direction of electrons
in various states. At low energies, the spins are completely in-plane.
They acquire a small out-of-plane component at higher energies. The
dotted lines represent incoming photons of helicity $-1$ (left-CP
photons). These photons can only \emph{raise} the $\langle S_{z}\rangle$
of an electron, and thus are preferentially absorbed by electrons
whose $\langle S_{z}\rangle<0$ in the valence band. The chemical
potential $\mu$ must be between the initial and final states for
any absorption to occur. \label{fig:absorption-imbalance} (b) Constant
energy contours for the surface conduction band of Bi$_{2}$Se$_{3}$.
Dark lines denote lower energy. (a) is drawn at $p_{y}=0$. (c) Geometry
of the experiment. Light is incident normally on (111) surface of
Bi$_{2}$Se$_{3}$. The dotted lines represent the mirror plane $m$
about which the lattice has a reflection symmetry. The current $j_{a2}(t)$
(see text) is along $\hat{x}$.}

\end{figure}

Historically, the Berry curvature has been associated with fascinating
phenomena such as the anomalous Hall effect\cite{HaldaneAHE} and
the integer quantum Hall effect\cite{TKNN} and therefore, it is exciting
that it appears in the response here. Its main implication here is
that is gives us a simple rule, in addition to the requirement of
the right symmetries, for identifying the perturbations that can give
a linear-in-time CPGE at normal incidence: we look for perturbations
that result in a non-zero Berry curvature. Put another way, we can
identify perturbations that have the right symmetries but still do
not give this current because the Berry curvature vanishes for these
perturbations. Importantly, for TI SSs, the requirement of a non-zero
Berry curvature amounts to the simple physical condition that the
spin-direction of the electrons have all three components non-zero.
In other words, if the electron spin in the SSs is completely in-plane,
the Berry curvature is zero and no linear-in-time CPGE is expected.
The spins must somehow be tipped slightly out of the plane, as shown
in Fig\ref{fig:absorption-imbalance}a, in order to get such a response.
Thus, a pure Dirac (linear) dispersion, for which the spins are planar,
cannot give this response; deviations from linearity, such as the
hexagonal warping on the (111) surface of Bi$_{2}$Te$_{3}$\cite{Fu3fold},
are essential for tilting the spins out of the plane.

\textcolor{black}{CPGE has been observed in the past in GaAs\cite{Ganichev GaAs}
and SiGe\cite{Ganichev SiGe} quantum wells - both systems with strong
spin-orbit coupling. However, no connection with the Berry curvature
was made in those cases. The connection, if present, may be harder
to find because the description there necessarily involves transitions
between four bands - two spin-orbit split valence bands and two spin-degenerate
conduction bands. In contrast, TI SSs can be faithfully treated within
a two-band model. Thus, TI SSs are a more convenient system theoretically
compared to semiconductor quantum wells. In general, if a surface
has no rotational symmetry about the surface normal, such a photocurrent
is allowed.}

Finally, we estimate the current on the (111) surface of Bi$_{2}$Se$_{3}$
using an effective model for the SSs\cite{Fu3fold,STImodel}. This
model captures the deviations from linearity of the SS dispersion
due to the threefold rotational symmetry of the (111) surface of Bi$_{2}$Se$_{3}$.
These deviations have been observed in photoemission experiments on
Bi$_{2}$Te$_{3}$\cite{Chen}. Similar deviations are expected for
Bi$_{2}$Se$_{3}$\cite{STImodel}, though they cannot be seen in
the slightly smaller momentum range compared to Bi$_{2}$Te$_{3}$
over which data is currently available\cite{BiSefermisurface}. In
order to get a direct current with CP light at normal incidence, rotational
symmetry about the surface normal needs to be broken. Based on the
requirement of non-zero Berry curvature, we propose to do this in
two ways:
\begin{enumerate}
\item by applying an in-plane magnetic field and including deviations from
linearity of the dispersion
\item by applying a strain.
\end{enumerate}
With a magnetic field of $10T$ (With a 1\% strain) and assuming a
scattering time of \textsf{\textcolor{black}{$10ps$}}\textcolor{blue}{,
}\textcolor{black}{(the scattering time in GaAs is $\sim1ns$ over
a wide range of temperatures\cite{GaAs relaxation time}; we use a
conservative estimate for Bi$_{2}$Se$_{3}$ here)} we find that a
current density of $\sim100nA/mm$ ($\sim10nA/mm$) can be obtained
due to the CPGE with a 1Watt laser. This value can be easily measured
by current experimental techniques. Conversely, the scattering time,
crucial for transport processes, for Bi$_{2}$Se$_{3}$ SSs can be
determined by measuring the current. In comparison, circular photogalvanic
currents of a few nanoamperes per Watt of laser power have been measured
in GaAs and SiGe quantum wells. 

A connection between the optical response of a system and the Berry
curvature of its bands has been previously noted at the low frequencies,
where a semiclassical mechanism involving the anomalous velocity of
electrons in a single band explains it\cite{DeyoPGE}. Here, we show
it for inter band transitions where no quasiclassical approximation
is applicable. Instead, we calculate the quadratic response function
directly. A connection is still present which points to a deeper relation
between the response functions and the Berry curvature.

This paper is organized as follows. In Sec. \ref{sec:Symmetry-considerations},
we state the symmetry conditions under which a CPGE may occur. We
present our results, both general as well as for Bi$_{2}$Se$_{3}$
in particular, in Sec. \ref{sub:Results} and describe the microscopic
mechanism in Sec. \ref{sub:Physical-process}. The calculation is
described briefly in Sec. \ref{sub:Calculation} and in detail in
Appendix \ref{sec:current calc}. In Sec. \ref{sec:spin generation},
we give our results for dc spin\textcolor{blue}{.}

\section{symmetry considerations for the CPGE\label{sec:Symmetry-considerations}}

In this section, we specify the symmetry conditions under which one
can get a CPGE on the surface of a TI. But first, let us briefly review
the concept of the CPGE in general.

The dominant dc response of matter to an oscillating electric field
is, in general, quadratic in the electric field. When the response
of interest is a current, the effect is known as the photogalvanic
effect. This current can be written as\begin{equation}
j_{\alpha}=\eta_{\alpha\beta\gamma}\mathcal{E}_{\beta}(\omega)\mathcal{E}_{\gamma}(-\omega)\end{equation}
where $\mathcal{E}_{\alpha}(t)=\mathcal{E}_{\alpha}(\omega)e^{i\omega t}+\mathcal{E}_{\alpha}^{*}(\omega)e^{-i\omega t}$
is the incident electric field, $\mathcal{E}_{\alpha}^{*}(\omega)=\mathcal{E}_{\alpha}(-\omega)$
and $\eta_{\alpha\beta\gamma}$ is a third rank tensor, which has
non-zero components only for systems that break inversion symmetry,
such as the surface of a crystal.

For $j_{\alpha}$ to be real, one has $\eta_{\alpha\beta\gamma}=\eta_{\alpha\gamma\beta}^{*}$.
Thus, the real (imaginary) part of $\eta_{\alpha\beta\gamma}$ is
symmetric (anti-symmetric) under interchange of $\beta$ and $\gamma$,
and therefore describes a current that is even (odd) under the transformation
$\omega\to-\omega$. Consequently, $j_{\alpha}$ can be conveniently
separated according to\begin{equation}
j_{\alpha}=\mathtt{S}_{\alpha\beta\gamma}\left(\frac{\mathcal{E}_{\beta}(\omega)\mathcal{E}_{\gamma}^{*}(\omega)+\mathcal{E}_{\beta}^{*}(\omega)\mathcal{E}_{\gamma}(\omega)}{2}\right)+i\mathtt{A}_{\alpha\mu}(\boldsymbol{\mathcal{E}}\times\boldsymbol{\mathcal{E}}^{*})_{\mu}\label{eq:intro current}\end{equation}
where $\mathtt{S}_{\alpha\beta\gamma}$ is the symmetric part of $\eta_{\alpha\beta\gamma}$
and $\mathtt{A}_{\alpha\mu}$ is a second-rank pseudo-tensor composed
of the anti-symmetric part of $\eta_{\alpha\beta\gamma}$. For CP
light, $\boldsymbol{\mathcal{E}}\propto\hat{x}\pm i\hat{y}$ if $\hat{z}$
is the propagation direction and only the second term in Eq. (\ref{eq:intro current})
survives, and hence represents the CPGE. This effect is odd in $\omega$.
On the other hand, the first term, which is even in $\omega$, represents
the linear photogalvanic effect as it is the only contribution for
linearly polarized light. Since the transformation $\omega\to-\omega$,
or equivalently, $\boldsymbol{\mathcal{E}}\to\boldsymbol{\mathcal{E}}^{*}$
reverses the helicity of CP light, i.e., changes right-CP light to
left-CP light and vice versa, the CPGE is the helicity-dependent part
of the photogalvanic effect.

The helicity of CP light is odd (i.e., right- and left-CP light get
interchanged) under time-reversal. It is also odd under mirror reflection
about a plane that contains the incident beam, but invariant under
arbitrary rotation about the direction of propagation. Let us consider
normal incidence of CP light on a TI surface normal to the $z$ axis.
Let us further assume that there is a mirror plane which is the $y$-$z$
plane (See Fig. \ref{fig:absorption-imbalance}c). Then the symmetries
above imply that the only component of direct current that reverses
direction on switching the helicity is a current along the $x$ axis.
If there is also rotation symmetry $R_{z}$ about the $z$-axis (such
as the threefold rotation symmetry on the (111) surface of Bi$_{2}$Se$_{3}$),
then no surface helicity-dependent direct photocurrent is permitted.
One needs to break this rotation symmetry completely by applying,
for example, and in-plane magnetic field, strain etc., to obtain a
nonvanishing current.

\section{helicity-dependent direct photocurrent}

We now present our main results for the photocurrent and estimate
it for Bi$_{2}$Se$_{3}$. After painting a simple microscopic picture
for the mechanism, we give a brief outline of the full quantum mechanical
treatment of the phenomenon.

\subsection{Results\label{sub:Results}}

A general two-band Hamiltonian (in the absence of the incident light)
can be written as

\begin{equation}
\mathbb{H}=\sum_{\mathbf{p}}H_{\mathbf{p}}=\sum_{\mathbf{p}}|E_{\mathbf{p}}|\mathbf{\hat{n}}(\mathbf{p}).\boldsymbol{\sigma}\label{eq:Hspecial}\end{equation}
upto a term proportional to the identity matrix, which is not important
for our main result which involves only inter-band transitions. Here
$\hat{\mathbf{n}}(\mathbf{p})$ is a unit vector and $\boldsymbol{\sigma}$
are the spin-Pauli matrices. Clearly, this can capture a Dirac dispersion,
eg. with $E(\mathbf{p})=\pm v_{F}p$ and $\hat{n}(\mathbf{p})=v_{F}\hat{\mathbf{z}}\times\mathbf{p}$.
It can also capture the SSs of Bi$_{2}$Se$_{3}$ in the vicinity
of the Dirac point, which includes deviations beyond the Dirac limit.
We also assume the Hamiltonian has a reflection symmetry $m$ about
$y$-axis, where $\hat{\mathbf{z}}$ is the surface normal. Using
the zero temperature quadratic response theory described in Sec \ref{sub:Calculation},
we calculate the current due to the CPGE and find that\begin{equation}
\vec{j}_{CPGE}(t)=\left(j_{na}+j_{a1}+j_{a2}(t)\right)\hat{\mathbf{x}}\label{eq:total current}\end{equation}
 where the subscripts $a$ ($na$) stand for {}``absorptive'' and
{}``non-absorptive'', respectively. The absorptive part of the response
involves a zero momentum interband transition between a pair of levels
separated by energy $\hbar\omega$. These terms are only non zero
when there is one occupied and one empty level. In this part of the
response, we find a term that is time-dependent, $j_{a2}(t)$. In
particular, this term grows linearly with the time over which the
electromagnetic perturbation is present, which is allowed for a dc
response. In reality, this linear growth is cut off by a decay process
which equilibrates populations, and is characterized by a time constant
$\tau$. In clean samples at sufficiently low temperatures, characterized
by large $\tau$, this contribution is expected to dominate the response,
and hence, is the focus of our work. The other contributions are discussed
in Appendix \ref{sec:current calc}. Conversely, because of the linear
growth with time, one can determine the lifetime of the excited states
by measuring the photocurrent. This term is \begin{equation}
j_{a2}(t)=-\frac{\pi e^{3}\hbar\mathcal{E}_{0}^{2}t\textrm{sgn}(\omega)}{4}\sum_{\mathbf{p}}\delta(\hbar|\omega|-2|E_{\mathbf{p}}|)v_{x}(\mathbf{p})F(\mathbf{p})\label{eq:current general}\end{equation}
 where we have assumed that the chemical potential is in between the
two energy levels $\pm|E_{\mathbf{p}}|$ connected by the optical
frequency $\hbar\omega$, and that temperature can be neglected compared
to this energy scale. Here, $v_{x}(\mathbf{p})=\frac{\partial|E_{\mathbf{p}}|}{\partial p_{x}}$
is the conventional velocity and $F(\mathbf{p})=i\sum_{\mathbf{p}}\langle\partial_{p_{x}}u(\mathbf{p})|\partial_{p_{y}}u(\mathbf{p})\rangle+c.c.$,
where $|u(\mathbf{p})\rangle$ is the conduction band Bloch state
at momentum $\mathbf{p}$, is the {\em Berry curvature} of the
conduction band at momentum $\mathbf{p}$. For the class of Hamiltonians
(\ref{eq:Hspecial}) that we are concerned with, the Berry curvature
is given by (See Appendix \ref{sec:Berry expression proof}): \begin{equation}
F(\mathbf{p})=\hat{\mathbf{n}}.\left(\frac{\partial\hat{\mathbf{n}}}{\partial p_{x}}\times\frac{\partial\hat{\mathbf{n}}}{\partial p_{y}}\right)\end{equation}
 which is the skyrmion density of the unit vector $\hat{\mathbf{n}}$
in momentum space. Since $\partial_{p_{i}}\hat{\mathbf{n}}\perp\hat{\mathbf{n}}$
for $i=x,y$, $F(\mathbf{p})\neq0$ only if all three components of
$\hat{\mathbf{n}}$ are nonvanishing. For linearly dispersing bands,
$\hat{\mathbf{n}}$ has only two non-zero components (eg. $H_{\mathbf{p}}=p_{y}\sigma_{x}-p_{x}\sigma_{y}$,
$\hat{\mathbf{n}}\propto(p_{y},-p_{x},0)$). Hence, corrections beyond
the pure Dirac dispersion are essential. Also, due to $m$, the Berry
curvature satisfies $F(p_{x},p_{y})=-F(-p_{x},p_{y})$. Since in Eq.
(\ref{eq:current general}) we have the $x$-velocity multiplying
the Berry curvature, which also transforms the same way, a finite
contribution is obtained on doing the momentum sum.

We now calculate $j_{a2}(t)$ for the threefold-symmetric (111) surface
of Bi$_{2}$Se$_{3}$ starting from the effective Hamiltonian\cite{Fu3fold,STImodel}\begin{equation}
H=v_{F}(p_{x}\sigma_{y}-p_{y}\sigma_{x})+\frac{\lambda}{2}\left(p_{+}^{3}+p_{-}^{3}\right)\sigma_{z}\label{eq:surface ham}\end{equation}
where $v_{F}\sim5\times10^{5}m/s$\cite{TIprediction} and $\lambda=50.1eV\cdot$Å$^{3}$\cite{STImodel}.
A spin independent quadratic term has been dropped since it does not
modify the answers for interband transitions, which only involve the
energy difference between the bands.

To get a non-zero $j_{CPGE}$, the threefold rotational symmetry must
be broken, which we first propose to do by applying a magnetic field
$B$ in the $x$-direction. This field has no orbital effect, and
can be treated by adding a Zeeman term $-g_{x}\mu_{B}B\sigma_{x}$,
where $g_{x}$ is the appropriate g-factor and $\mu_{B}$ is the Bohr
magneton, to the Hamiltonian (\ref{eq:surface ham}). To lowest order
in $\lambda$ and $B$, we get\begin{equation}
j_{a2}(t)=\frac{3e^{3}v_{F}\mathcal{E}_{0}^{2}\lambda(g_{x}\mu_{B}B)^{2}t}{16\hbar^{2}\omega}\mathcal{A}\label{eq:current with B}\end{equation}
to lowest order in $\lambda$ and $B^{2}$, where $\mathcal{A}$ is
the laser spot-size. For $g_{x}=0.5$\cite{STImodel}, and assuming
the experiment is done in a $10T$ field with a continuous wave laser
with $\hbar\omega=0.1eV$ which is less than the bulk band gap of
$0.35eV$\cite{Xia}, $\mathcal{A}\sim1mm^{2}$, a laser power of
$1W$, and the spin relaxation time \textsf{\textcolor{black}{$t\sim10ps$}},
we get a current density of \textsf{\textcolor{black}{$\sim100nA/mm$}},
which is easily measurable by current experimental techniques. Note
that the expression (\ref{eq:current with B}) for $j_{a2}(t)$ contains
the parameter $\lambda$ which measures the coupling to $\sigma_{z}$
in Eq. (\ref{eq:surface ham}). Since $\vec{B}=B\hat{\mathbf{x}}$
breaks the rotation symmetry of the surface completely, a naive symmetry
analysis suggests, wrongly, that deviations from linearity, measured
by $\lambda$, are not needed to get $j_{a2}(t)$.

The rotation symmetry can also be broken by applying a strain along
$x$, which can be modeled by adding a term $\delta\lambda p_{x}^{3}\sigma_{z}$
to $H$ in Eq.\ref{eq:surface ham}). This gives \begin{equation}
j_{a2}(t)=\frac{3e^{3}v_{F}(\delta\lambda)\mathcal{E}_{0}^{2}\omega t}{2^{7}}\mathcal{A}\label{eq:current with strain}\end{equation}
to lowest order in $\lambda$ and $\delta\lambda$. For a 1\% strain,
$\delta\lambda/\lambda=0.01$, and the same values for the other paramaters
as in Eq.(\ref{eq:current with B}), we get a current density of \textsf{\textcolor{black}{$\sim10nA/mm$}}.
Eq. (\ref{eq:current with strain}) does not contain $\lambda$; this
is because $\delta\lambda$ alone both breaks the rotation symmetry
and tips the spins out of the $xy$-plane.

\subsection{Physical process\label{sub:Physical-process}}

The appearance of the Berry curvature suggests a role of the anomalous
velocity in generating the current. Such mechanisms have been discussed
in the literature in the context of the CPGE\cite{DeyoPGE,Moore_Orenstein_Berry_curvature}.
However, those mechanisms only work when the electric field changes
slowly compared to the typical scattering time. The SSs of Bi$_{2}$Se$_{3}$
probably have lifetimes of tens of picoseconds, and thus, we are in
the opposite limit when $\hbar\omega=0.1eV$, which corresponds to
a time scale \textsf{\textcolor{black}{$10^{3}$}} times shorter.

In this limit, the dc responses are a result of a preferential absorption
of the photon at one of the two momentum points for each pair of points
$(\pm p_{x},p_{y})$ related by $m$, as shown in Fig. \ref{fig:absorption-imbalance}a
for $p_{y}=0$. According to the surface Hamiltonian (\ref{eq:surface ham}),
the spin vector $\mathbf{S}=\frac{\boldsymbol{\sigma}}{2}\hbar$ gets
tipped out of the $xy$-plane for states that lie beyond the linear
dispersion regime, but the direction of the tipping is opposite for
$(p_{x},p_{y})$ and $(-p_{x},p_{y})$. Thus, photons of helicity
$-1$, which can only \emph{raise} $\langle S_{z}\rangle$ of an electron,
are preferentially absorbed by the electrons that have $\langle S_{z}\rangle<0$
in the ground state. The response, then, is determined by the properties
of these electrons. Clearly, the process is helicity-dependent as
reversing the helicity would cause electrons with $\langle S_{z}\rangle>0$
to absorb the light preferentially.

This is consistent with the requirement of a non-zero Berry curvature,
which essentially amounts to the spin direction $\hat{\mathbf{n}}$
having to be a three-dimensional vector. In the linear limit, where
$H=v_{F}(p_{x}\sigma_{y}-p_{y}\sigma_{x})$, the spin is entirely
in-plane, and all the electrons absorb the incident light equally.

\subsection{Calculation in brief\label{sub:Calculation}}

We now briefly outline the calculation of the helicity-dependent photocurrent.
The detailed calculation can be found in Appendix \ref{sec:current calc}.
Readers only interested in our results may wish to skip this section.

\textbf{The Model:} The Hamiltonian and relevant electric field (vector
potential) perturbations for getting a direct current to second order
in the electric field of the incident photon are\begin{eqnarray}
H & = & |E_{\mathbf{p}}|\hat{\mathbf{n}}(p).\boldsymbol{\sigma}\label{eq:hamiltonian}\\
H^{\prime} & = & j_{x}A_{x}(t)+j_{y}A_{y}(t)\label{eq:perturbation}\\
j_{\alpha} & = & \frac{\partial H}{\partial p_{\alpha}}\label{eq:current operator}\\
A_{x}(t)+iA_{y}(t) & = & A_{0}e^{i(\omega-i\epsilon)t}\label{eq:vector potential}\end{eqnarray}
 where $\mathbf{A}$ is the vector potential, $\hat{z}$ is assumed
to be the surface normal, and $\epsilon$ is a small positive number
which ensures slow switch-on of the light.

\textbf{Quadratic response Theory:} In general, the current along
$x$ to all orders in the perturbation $H^{\prime}$ is \begin{equation}
\langle j_{x}\rangle(t)=\left\langle T^{*}\left(e^{i\int_{-\infty}^{t}dt^{\prime}H^{\prime}(t^{\prime})}\right)j_{x}(t)T\left(e^{-i\int_{-\infty}^{t}dt^{\prime}H^{\prime}(t^{\prime})}\right)\right\rangle \label{eq:all orders of perturbation}\end{equation}
 where $T\,(T^{*})$ denotes time-ordering (anti-time-ordering) and
$O(t)=e^{iHt}Oe^{-iHt}$. Terms first order in $H^{\prime}$ cannot
give a direct current. The contribution to the current from the second
order terms can be written as \begin{multline}
\langle j_{x}\rangle(t)=\intop_{-\infty}^{t}dt^{\prime}\intop_{-\infty}^{t_{1}}dt^{\prime\prime}\left\langle \left[\left[j_{x}(t),H^{\prime}(t^{\prime})\right],H^{\prime}(t^{\prime\prime})\right]\right\rangle \\
=\intop_{-\infty}^{t}dt^{\prime}\intop_{-\infty}^{t_{1}}dt^{\prime\prime}\chi_{x\alpha\beta}(t,t^{\prime},t^{\prime\prime})A_{\alpha}(t^{\prime})A_{\beta}(t^{\prime\prime})\end{multline}
where $\alpha,\beta\in\{x,y\}$, $\chi_{x\alpha\beta}(t,t^{\prime},t^{\prime\prime})=\chi_{x\alpha\beta}(0,t^{\prime}-t,t^{\prime\prime}-t)=\left\langle \left[\left[j_{x},j_{\alpha}(t^{\prime}-t)\right],j_{\beta}(t^{\prime\prime}-t)\right]\right\rangle \equiv\chi_{x\alpha\beta}(t^{\prime}-t,t^{\prime\prime}-t)$
due to time translational invariance, and the expectation value is
over the ground state which has all states with $E_{\mathbf{p}}<(>)\,0$
filled (empty). For Hamiltonians of the form of Eq. (\ref{eq:hamiltonian}),
the expectation value of any traceless operator $O$ in the Fermi
sea ground state can be written as a trace:\begin{equation}
\langle O\rangle=\sum_{\mathbf{p}}\frac{1}{2}\mathrm{Tr}\left\{ \left(1-\frac{H}{|E_{\mathbf{p}}|}\right)O\right\} =-\sum_{\mathbf{p}}\frac{\mathrm{Tr}\left(HO\right)}{2|E_{\mathbf{p}}|}\label{eq:avg_to_trace}\end{equation}
 This gives, \begin{align}
\chi_{x\alpha\beta}(t_{1},t_{2}) & =-\sum_{p}\frac{\mathrm{Tr}\left(H\left[\left[j_{x},j_{\alpha}(t_{1})\right],j_{\beta}(t_{2})\right]\right)}{2|E_{\mathbf{p}}|}\label{eq:chi_definition}\end{align}
 Eq. (\ref{eq:chi_definition}) is the zero temperature limit of the
finite temperature expression for the quadratic susceptibility proven
in Ref. \cite{NLO_Butcher}.

Because of the mirror symmetry $m$, $\chi_{x\alpha\beta}(t_{1},t_{2})$
is non-vanishing only for $\alpha\neq\beta$. To get a direct current,
we retain only the non-oscillating part of $A_{x}(t+t_{i})A_{y}(t+t_{j})=\frac{A_{0}^{2}}{2}e^{2\epsilon t}\left[\sin\left(2\omega t+\omega(t_{i}+t_{j})\right)-\sin\left(\omega(t_{i}-t_{j})\right)\right]$.
Thus,\begin{multline}
j_{x}^{dc}(t)=\frac{A_{0}^{2}e^{2\epsilon t}}{4}\intop_{-\infty}^{0}dt_{1}\intop_{-\infty}^{t_{1}}dt_{2}\bigg\{\left(\chi_{xxy}-\chi_{xyx}\right)(t_{1},t_{2})\times\\
e^{\epsilon(t_{1}+t_{2})}\sin\left(\omega(t_{2}-t_{1})\right)\bigg\}\label{eq:jxdc}\end{multline}

\textbf{The Result:} After carrying out the two time-integrals, we
get the three currents mentioned in Eq. (\ref{eq:total current}).
For clean samples at low temperatures, $j_{a2}(t)$, which grows linearly
with time, is expected to dominate. A general expression for this
term is (in the units $e=\hbar=v_{F}=1$ where $v_{F}$ is the Fermi
velocity)\begin{align}
j_{a2}(t) & =\nonumber \\
 & \frac{iA_{0}^{2}\pi t\text{sgn}(\omega)}{2\omega^{2}}\sum_{\mathbf{p}}\delta(|\omega|-2|E_{\mathbf{p}}|)\mathrm{Tr}(Hj_{x})\mathrm{Tr}(H[j_{x},j_{y}])\label{eq:current as trace}\end{align}
 Using Eqs. (\ref{eq:hamiltonian}) and (\ref{eq:current operator})
and the Lie algebra of the Pauli matrices, $[\sigma_{i},\sigma_{j}]=2i\epsilon_{ijk}\sigma_{k}$
where $\epsilon_{ijk}$ is the anti-symmetric tensor, the above traces
can be written as\begin{eqnarray}
\mathrm{Tr}(Hj_{x}) & = & 2|E_{\mathbf{p}}|v_{x}(\mathbf{p})\label{eq:velocity as trace}\\
\mathrm{Tr}(H\left[j_{x},j_{y}\right]) & = & 4i|E_{\mathbf{p}}|^{3}\hat{\mathbf{n}}.\left(\frac{\partial\hat{\mathbf{n}}}{\partial p_{x}}\times\frac{\partial\hat{\mathbf{n}}}{\partial p_{y}}\right)\nonumber \\
 & = & 4i|E_{\mathbf{p}}|^{3}F(\mathbf{p})\label{eq: curvature}\end{eqnarray}
 Eqs. (\ref{eq:current as trace}), (\ref{eq:velocity as trace})
and (\ref{eq: curvature}) give our main result Eq. (\ref{eq:current general}).

\section{spin generation\label{sec:spin generation}}

Having understood the microscopic mechanism underlying the generation
of the photocurrent $j_{a2}(t)$ , we wonder, next, whether such a
population imbalance can lead to any other helicity-dependent macroscopic
responses. Since each absorbed photon flips the $z$-component of
the spin of an electron, a net $\langle S_{z}\rangle$ is expected
to be generated on the surface.

The calculation of $\langle S_{z}\rangle$ is identical to that of
$j_{CPGE}$. The total $\langle S_{z}\rangle$ generated consists
of the same three parts as $j_{CPGE}$, and the dominant part is\begin{equation}
S_{a2}^{z}(t)=-\frac{\pi e^{2}\mathcal{E}_{0}^{2}\hbar t\mbox{sgn}(\omega)}{8}\sum_{\mathbf{p}}\delta(\hbar|\omega|-2|E_{p}|)n_{z}(\mathbf{p})F(\mathbf{p})\label{eq:Sz general}\end{equation}
$S_{z}$ does not break the rotational symmetry of the surface, so
we calculate $S_{a2}^{z}(t)$ directly for the threefold symmetric
Hamiltonian (\ref{eq:surface ham}) and obtain\begin{equation}
S_{a2}^{z}(t)=\frac{e^{2}\mathcal{E}_{0}^{2}(\hbar\omega)^{3}\lambda^{2}t}{2^{10}}\mathcal{A}\label{eq:spin result}\end{equation}
For the same values of all the parameters as for $j_{a2}(t)$, we
get \textsf{\textcolor{black}{$S_{a2}^{z}(t)\sim10\hbar$}}, which
means only ten electron spins are flipped over an area of $\sim1mm^{2}$.
If, instead, we ignore the cubic corrections but assume magnetic ordering
on the surface, so that $H=v_{F}(p_{x}\sigma_{y}-p_{y}\sigma_{x})+M\sigma_{z}$,
we get \begin{equation}
S_{a2}^{z}(t)=-\frac{e^{2}\mathcal{E}_{0}^{2}M^{2}t}{16(\hbar\omega)^{3}}\mathcal{A}\end{equation}
which again gives a rather small value of \textsf{\textcolor{black}{$\sim10\hbar$}}
for $M\sim10K$, a typical magnetic ordering transition temperature.
However, the spin generated could be measurable if one uses a pulsed
laser of, say, MegaWatt power, and performs a time-resolved experiment.

\section{Conclusions}

In summary, we studied the CPGE on the surface of a TI at normal incidence,
and applied the results to the (111) surface of Bi$_{2}$Se$_{3}$.
If the rotational symmetry of the TI surface is broken by applying
an in-plane magnetic field or a strain, we predict an experimentally
measurable direct photocurrent. A striking feature of this current
is that it depends on the Berry curvature of the electron bands. Such
a dependence can be understood intuitively as a result of the incident
photons getting absorbed unequally by electrons of different momenta
and hence, different average spins. The current grows linearly with
time until a decay process equilibrates populations, which provides
a way of determining the excited states lifetime. We also calculated
the amount of dc helicity-dependent out-of-plane component of the
electron spin generated. This does not require any rotational symmetry
breaking; however, the numerical value is rather small with typical
values of parameters.

In the future, we hope to find a generalization of our results for
oblique incidence. Experimentally, this is a very attractive way of
breaking the rotational symmetry of the surface; indeed, such experiments
have already been performed successfully on graphene\cite{graphene photocurrent}.
In graphene, helicity-dependent direct photocurrents have also been
predicted by applying a dc bias\cite{PhotoHallEffect}. However, with
a dc bias across a TI surface and ordinary continuous lasers, we find
the current to be too low to be measurable. Finally, we also wonder
whether the Berry curvature dependence of the helicity-dependent response
to CP light survives for three- and higher-band models. If it does,
it would be interesting to write such a model for semiconductor quantum
wells such as GaAs and SiGe. It could also enable one to treat oblique
incidence, by considering transitions to higher bands of different
parities, because they are driven by the normal component of the electric
field, $E_{z}$.\\

We would like to thank Ashvin Vishwanath for enlightening discussions,
Joseph Orenstein for useful experimental inputs, and Ashvin Vishwanath
and Yi Zhang for invaluable feedback on the draft.

This work was supported by LBNL DOE-504108.

\appendix

\section{Proof of Berry curvature expression\label{sec:Berry expression proof}}

Here we show that the Berry curvature defined for Bloch electrons
as \begin{equation}
F(\mathbf{p})=i\left(\langle\partial_{p_{x}}u|\partial_{p_{y}}u\rangle-\langle\partial_{p_{y}}u|\partial_{p_{x}}u\rangle\right){\color{blue}}\label{eq:curvature bloch}\end{equation}
 can be written as \begin{equation}
F(\mathbf{p})=\hat{\mathbf{n}}.\left(\partial_{p_{x}}\hat{\mathbf{n}}\times\partial_{p_{y}}\hat{\mathbf{n}}\right)\label{eq:curvature unit vector}\end{equation}
 for the band with energy $|E_{\mathbf{p}}|$ for Hamiltonians of
the form $H_{\mathbf{p}}=|E_{\mathbf{p}}|\hat{\mathbf{n}}(\mathbf{p}).\boldsymbol{\sigma}$.

At momentum $\mathbf{p}$, the Bloch state $|u_{\mathbf{p}}\rangle$
with energy $|E_{\mathbf{p}}|$ is defined as the state whose spin
is along $\hat{\mathbf{n}}(\mathbf{p})$. Defining $|\uparrow\rangle$
as the state whose spin is along $+\hat{\mathbf{z}}$, $|u_{\mathbf{p}}\rangle$
is obtained by performing the appropriate rotations,\begin{equation}
|u_{\mathbf{p}}\rangle=e^{-i\frac{\sigma_{z}}{2}\phi(\mathbf{p})}e^{i\frac{\sigma_{y}}{2}\theta(\mathbf{p})}|\uparrow\rangle\label{eq:state at p}\end{equation}
where $\theta(\mathbf{p})$ and $\phi(\mathbf{p})$ are the polar
angles that define $\hat{\mathbf{n}}(\mathbf{p})$:\begin{equation}
\hat{\mathbf{n}}(\mathbf{p})=\sin\theta(\mathbf{p})\cos\phi(\mathbf{p})\hat{x}+\sin\theta(\mathbf{p})\sin\phi(\mathbf{p})\hat{y}+\cos\theta(\mathbf{p})\hat{z}\label{eq:n polar}\end{equation}

Substituting Eq. (\ref{eq:state at p}) in Eq. (\ref{eq:curvature bloch}),
one gets\begin{equation}
F(\mathbf{p})=\sin\theta(\mathbf{p})\left(\partial_{p_{x}}\theta(\mathbf{p})\partial_{p_{y}}\phi(\mathbf{p})-\partial_{p_{x}}\phi(\mathbf{p})\partial_{p_{y}}\theta(\mathbf{p})\right)\end{equation}
which, on using Eq. (\ref{eq:n polar}) and some algebra, reduces
to the required expression Eq. (\ref{eq:curvature unit vector}).

\section{current calculation for the cpge\label{sec:current calc}}

Here we explain the current-calculation of Sec. \ref{sub:Results}
in more detail and also state results for the parts of the current
that we chose not to focus on there.

As shown in Sec. \ref{sub:Calculation}, the relevant susceptibility
is\begin{eqnarray}
\chi^{x\alpha\beta}(t,t^{\prime},t^{\prime\prime}) & = & -\frac{1}{2}\sum_{\mathbf{p}}\mathrm{Tr}\left(\frac{H}{|E_{\mathbf{p}}|}\left[\left[j^{x}(t),j^{\alpha}(t^{\prime})\right],j^{\beta}(t^{\prime\prime})\right]\right)\nonumber \\
 & = & -\sum_{\mathbf{p}}\frac{1}{2|E_{\mathbf{p}}|}\mathrm{Tr}\left(H\left[\left[j^{x},j^{\alpha}(t_{1})\right],j^{\beta}(t_{2})\right]\right)\nonumber \\
 & \equiv & \chi^{x\alpha\beta}(t_{1},t_{2})\label{eq:chi_defintion}\end{eqnarray}
 where $t_{1}=t^{\prime}-t,\, t_{2}=t^{\prime\prime}-t$, and the
non-vanishing components of $\chi^{x\alpha\beta}$ are those for which
$\alpha\neq\beta$. The non-oscillating part of the current, hence,
is\begin{multline}
\langle j_{x}^{dc}\rangle(t)=j_{CPGE}(t)=\frac{A_{0}^{2}e^{2\epsilon t}}{4}\intop_{-\infty}^{0}dt_{1}\intop_{-\infty}^{t_{1}}dt_{2}\\
\left(\chi^{xxy}(t_{1},t_{2})-\chi^{xyx}(t_{1},t_{2})\right)e^{\epsilon(t_{1}+t_{2})}\sin\left(\omega(t_{2}-t_{1})\right)\end{multline}
Since $j_{CPGE}(t)$ is an odd function of $\omega$, it reverses
on reversing the polarization, as expected.

The traces in the susceptibility expressions are calculated by introducing
a complete set of states in place of the identity several times. Thus,\begin{align}
 & \chi^{xxy}(t_{1},t_{2})\label{eq:chi 1}\\
= & -\sum_{\mathbf{p}}\frac{1}{2|E_{\mathbf{p}}|}\mathrm{Tr}\left(H\left[\left[j^{x},j^{x}(t_{1})\right],j^{y}(t_{2})\right]\right)\nonumber \\
= & -\frac{1}{2}\sum_{\mathbf{p}}\sum_{nml}\textrm{sgn}(E_{n})\biggl\{ e^{i(E_{m}-E_{n})t_{2}}\times\nonumber \\
 & \left(e^{i(E_{l}-E_{m})t_{1}}-e^{-i(E_{l}-E_{n})t_{1}}\right)X_{nl}X_{lm}Y_{mn}+\textrm{c.c.}\biggl\}\nonumber \end{align}
 where $X_{nl}=\langle n\left|j_{x}\right|m\rangle$ etc. and the
subscript $\mathbf{p}$ on $E_{\mathbf{p}}$ has been dropped to enhance
the readability. Similarly,\begin{align}
 & \chi^{xyx}(t_{1},t_{2})\label{eq:chi 2}\\
 & =-\sum_{\mathbf{p}}\frac{1}{2E_{\mathbf{p}}}\mathrm{Tr}\left(H\left[\left[j^{x},j^{y}(t_{1})\right],j^{x}(t_{2})\right]\right)\nonumber \\
 & =-\frac{1}{2}\sum_{\mathbf{p}}\sum_{nml}\textrm{sgn}(E_{n})\biggl\{ e^{i(E_{m}-E_{n})t_{2}}X_{mn}\times\nonumber \\
 & \left(e^{i(E_{l}-E_{m})t_{1}}X_{nl}Y_{lm}-e^{-i(E_{l}-E_{n})t_{1}}Y_{nl}X_{lm}\right)+\textrm{c.c.}\biggl\}\nonumber \end{align}

Substituting (\ref{eq:chi 1}) and (\ref{eq:chi 2}) in (\ref{eq:DC current expression}),
we get \begin{align}
 & j_{CPGE}(t)=\frac{A_{0}^{2}e^{2\epsilon t}}{4}\mathfrak{Re}\intop_{-\infty}^{0}dt_{1}\intop_{-\infty}^{t_{1}}dt_{2}e^{\epsilon(t_{1}+t_{2})}\times\\
 & \sin\left(\omega(t_{1}-t_{2})\right)\sum_{\mathbf{p},nml}\textrm{sgn}(E_{n})e^{i(E_{m}-E_{n})t_{2}}\times\nonumber \\
 & \biggl\{\left(e^{i(E_{l}-E_{m})t_{1}}-e^{-i(E_{l}-E_{n})t_{1}}\right)X_{nl}X_{lm}Y_{mn}-\nonumber \\
 & X_{mn}\left(e^{i(E_{l}-E_{m})t_{1}}X_{nl}Y_{lm}-e^{-i(E_{l}-E_{n})t_{1}}Y_{nl}X_{lm}\right)\biggl\}\nonumber \end{align}
 where $\mathfrak{Re}$ stands for `the real part of'. Carrying out
the the two time integrations gives\begin{align}
 & j_{CPGE}(t)=\frac{A_{0}^{2}e^{2\epsilon t}}{8}\mathfrak{Im}\sum_{\mathbf{p}}\sum_{nml}\textrm{sgn}(E_{n})\times\\
 & \left[\frac{1}{E_{m}-E_{n}+\omega-i\epsilon}-\frac{1}{E_{m}-E_{n}-\omega-i\epsilon}\right]\times\nonumber \\
 & \left\{ \frac{X_{nl}\left(X_{lm}Y_{mn}-Y_{lm}X_{mn}\right)}{E_{l}-E_{n}-2i\epsilon}+\frac{X_{lm}\left(Y_{mn}X_{nl}-X_{mn}Y_{nl}\right)}{E_{l}-E_{m}+2i\epsilon}\right\} \nonumber \end{align}
 where $\mathfrak{Im}$ stands for `the imaginary part of'. Using
$\mathfrak{Im}\left(\frac{1}{\Omega-i\epsilon}\right)=\pi\delta(\Omega)$
and $\mathfrak{Re}\left(\frac{1}{\Omega-i\epsilon}\right)=\frac{1}{\Omega}$
in the limit $\epsilon\to0$, we get after some algebra, $j_{CPGE}(t)=j_{na}+j_{a1}+j_{a2}(t)$,
where ($\mathrm{Tr}$ denotes the trace)\begin{gather}
j_{na}=\frac{A_{0}^{2}}{16}\sum_{\mathbf{p}}\frac{\omega(\omega^{2}-12E_{\mathbf{p}}^{2})}{i|E_{\mathbf{p}}|^{3}(\omega^{2}-4E_{\mathbf{p}}^{2})^{2}}\times\nonumber \\
\mathrm{Tr}(Hj_{x})\mathrm{Tr}(H\left[j_{x},j_{y}\right])\end{gather}
 comes from intraband processes and is constant in time,\begin{gather}
j_{a1}=-\frac{\pi A_{0}^{2}\textrm{sgn}(\omega)}{32}\sum_{\mathbf{p}}\frac{\delta(|\omega|-2|E_{\mathbf{p}}|)}{E_{\mathbf{p}}^{2}}\times\nonumber \\
\mathrm{Tr}(H\left[j_{x},\left[j_{x},j_{y}\right]\right])\end{gather}
 is a result of an interband transition absorption as indicated by
the $\delta$-function in energy and is also constant in time, and
\begin{gather}
j_{a2}(t)=i\frac{A_{0}^{2}\pi t\,\textrm{sgn}(\omega)}{8}\sum_{p}\delta(|\omega|-2|E_{\mathbf{p}}|)\times\nonumber \\
\frac{\mathrm{Tr}(Hj_{x})\mathrm{Tr}(H\left[j_{x},j_{y}\right])}{E_{\mathbf{p}}^{2}}\end{gather}
 which also results from interband absorption and increases linearly
in time. The last term was the main focus of our work.

\end{document}